\documentclass[twocolumn]{aastex63}

\received{2020 September 15 }
\revised{2020 December 7 }
\accepted{2020 December 21}
\submitjournal{ApJ}

\shorttitle{Temperature in solar sources of $^3$He-rich solar energetic particles}
\shortauthors{Bu\v{c}\'ik et al.}

\begin{document}

\title{Temperature in Solar Sources of $^3$He-rich Solar Energetic Particles and Relation to Ion Abundances}

\correspondingauthor{Radoslav Bu\v{c}\'ik}
\email{radoslav.bucik@swri.org}

\author{Radoslav Bu\v{c}\'ik}
\affiliation{Southwest Research Institute, San Antonio, TX 78238, USA}

\author{Sargam M. Mulay}
\affiliation{School of Physics and Astronomy, University of Glasgow, Glasgow, G12 8QQ, UK}

\author{Glenn M. Mason}
\affiliation{Applied Physics Laboratory, Johns Hopkins University, Laurel, MD 20723, USA}

\author{Nariaki V. Nitta}
\affiliation{Lockheed Martin Advanced Technology Center, Palo Alto, CA 94304, USA}

\author{Mihir I. Desai}
\affiliation{Southwest Research Institute, San Antonio, TX 78238, USA}
\affiliation{Department of Physics and Astronomy, University of Texas at San Antonio, San Antonio, TX 78249, USA}

\author{Maher A. Dayeh}
\affiliation{Southwest Research Institute, San Antonio, TX 78238, USA}
\affiliation{Department of Physics and Astronomy, University of Texas at San Antonio, San Antonio, TX 78249, USA}



\begin{abstract}

$^3$He-rich solar energetic particles (SEPs) are believed to be accelerated in solar flares or jets by a mechanism that depends on the ion charge-to-mass ($Q/M$) ratio. It implies that the flare plasma characteristics (e.g., temperature) may be effective in determining the elemental abundances of $^3$He-rich SEPs. This study examines the relation between the suprathermal ($\lesssim$0.2\,MeV\,nucleon$^{-1}$) abundances of the He--Fe ions measured on the  {\sl Advanced Composition Explorer} and temperature in the solar sources for 24 $^3$He-rich SEP events in the period 2010--2015. The differential emission measure technique is applied to derive the temperature of the source regions from the extreme ultraviolet imaging observations on the {\sl Solar Dynamics Observatory}. The obtained temperature distribution peaks at 2.0--2.5\,MK that is surprisingly consistent with earlier findings based on in-situ elemental abundance or charge state measurements. We have found a significant anti-correlation between $^3$He/$^4$He ratio and solar source temperature with a coefficient $-$0.6. It is most likely caused by non-charge-stripping processes, as both isotopes would be fully ionized in the inferred temperature range. This study shows that the elemental ratios $^4$He/O, N/O, Ne/O, Si/O, S/O, Ca/O, Fe/O generally behave with temperature as expected from abundance enhancement calculations at ionization equilibrium. The C and Mg, the two species with small changes in the $Q$/$M$ ratio in the obtained temperature range, show no such behavior with temperature and could be influenced by similar processes as for the $^3$He/$^4$He ratio.  

\end{abstract}



\section{Introduction} \label{sec:intro}

$^3$He-rich solar energetic particles (SEPs) are characterized by enormous enhancements of rare species like $^3$He or ultra-heavy ions (mass $>$ 70\,AMU) by factors up to 10$^4$ above the coronal or solar wind abundances \citep[see, e.g.,][for a review]{1984SSRv...38...89K,2007SSRv..130..231M,2013SSRv..175...53R,2017LNP...932.....R,2018SSRv..214...61R}. In a typical $^3$He-rich SEP event, $^4$He, $^{12}$C, $^{14}$N, $^{16}$O are unenhanced, while heavier species like $^{20}$Ne, $^{24}$Mg, $^{28}$Si, show enhancement that increases with the ion mass \citep{1994ApJS...90..649R,2004ApJ...606..555M}. The abundance of the light $^3$He isotope is in contrast with such enhancement pattern. Whereas, with He, the abundance of the lighter isotope is enhanced, with heavy ions, the heavier isotopes are enhanced \citep{1994ApJ...425..843M,2010ApJ...719.1212W}.

Solar sources of $^3$He-rich SEPs have been associated with extreme ultraviolet (EUV) coronal jets \citep[see][and references therein]{2020SSRv..216...24B}, indicating acceleration in magnetic reconnection region involving field lines open to interplanetary (IP) space \citep{2002ApJ...571L..63R,2006ApJ...639..495W}. The events show a high association ($\sim$99\%) with type-III radio bursts \citep[e.g.,][]{1986ApJ...308..902R,2006ApJ...650..438N}, the emission generated by low-energy electrons \citep[$<$30\,keV;][]{1999ApJ...519..864K} escaping into IP space. Furthermore, $^3$He-rich SEP events have been accompanied by minor X-ray flares \citep[mostly B- and C- class; e.g.,][]{2006ApJ...650..438N,2015ApJ...806..235N,2009ApJ...700L..56M,2016ApJ...833...63B}. 

Early measurements of high ionic charge states \citep{1987ApJ...317..951L} indicated $>$10\,MK source temperature of $^3$He-rich SEPs. However, an increase of charge states with kinetic energy \citep[e.g.,][]{2008ApJ...687..623D} has suggested that the mechanism responsible for high charge states is an additional ionization in a dense environment after/during acceleration \citep[e.g.,][]{1994ApJS...90..649R,2007SSRv..130..273K}. \citet{1994ApJS...90..649R} examined 228 $^3$He-rich events in August 1978 -- April 1991 with $^3$He/$^4$He $>$0.1 at the energy range of 1.2--1.6\,MeV\,nucleon$^{-1}$. Their abundance enhancement pattern has suggested that C, N, O are fully ionized as $^4$He, but Ne, Mg, Si are only partially stripped of their orbital electrons. The data on equilibrium ionization states in astrophysical plasma shows that this can happen in a temperature range of $\sim$3--5\,MK \citep{1994ApJS...90..649R}. In a more recent survey, \citet{2014SoPh..289.3817R,2014SoPh..289.4675R,2015SoPh..290.1761R} examined 111 Fe-rich impulsive events in April 1995 -- April 2013 at the energy range 2--10\,MeV\,nucleon$^{-1}$, and reported the most probable temperature range of $\sim$2.5--3.2\,MK. Note that several events have been reported with enhanced abundances of C and N, suggesting temperatures $\sim$0.5--1.5\,MK \citep{2002ApJ...565L..51M,2016ApJ...823..138M}. These temperatures may indicate that ions are accelerated very early \citep{2016ApJ...823..138M} and (or) on open field lines where heating is minimal \citep{2015SoPh..290.1761R}. The required low temperatures may also be due to a cool material observed in the form of a mini-filament in some $^3$He-rich SEP sources \citep{2016ApJ...823..138M,2016AN....337.1024I,2018ApJ...852...76B}.

At low energies, below $\sim$0.2\,MeV\,nucleon$^{-1}$, both equilibrium and non-equilibrium charge states depend only on temperature \citep[e.g.,][]{2000A&A...357..716K,2001A&A...375.1075K}  and not on charge stripping effects causing the energy dependence of heavy-ion charge states observed in $^3$He-rich events at higher energies, $\sim$0.2--1\,MeV\,nucleon$^{-1}$ \citep{2007SSRv..130..273K,2008ApJ...687..623D}. Therefore, the source temperature in $^3$He-rich SEP events can be determined from measurements of charge states at low energies. \citet{2008ApJ...687..623D} have obtained temperatures in the range of 1.3 to 3\,MK using low-energy (0.062--0.110\,MeV\,nucleon$^{-1}$) Fe charge states measurements in 14 $^3$He-rich SEP events in May 1998 -- September 2000.

The ionic charge states were inferred with help of measured isotopic ratio enhancements such as $^{22}$Ne/$^{20}$Ne \citep{1999GeoRL..26.2697C,1999GeoRL..26..149C,2010ApJ...719.1212W}. \citet{2010ApJ...719.1212W} have reported a corresponding source temperature of $\sim$4\,MK for a single $^3$He-rich SEP event and \citet{1999GeoRL..26.2697C} have reported the same temperature for nine large SEP events that were heavy-ion rich but show small $^3$He abundance ($^3$He/$^4$He $<$0.01 in most events).

\citet{1988ApJ...325L..53R} reported an increase of 2--3\,MeV\,nucleon$^{-1}$ heavy ion (C, Mg, Si, S, Fe) enhancements (relative to $^4$He) with soft X-ray temperature in the range 10--16\,MK. The origin of the reported dependencies remains unclear since many of these species are fully stripped at these temperatures \citep{1988ApJ...325L..53R}. \citet{1990ApJS...73..235R} has suggested that plasma waves escaping from heated closed regions accelerate nearby $^3$He-rich SEPs, possibly causing an increase of heavy-ion enhancements. The 1.3--1.6\,MeV\,nucleon$^{-1}$ $^3$He/$^4$He ratio showed no correlation with soft X-ray temperature \citep{1988ApJ...327..998R}. The temperature was determined from ISEE soft X-ray measurements at the time of hard X-ray maximum or at the time of type-III radio burst (at 2\,MHz that corresponds to $\sim$6\,R$_\sun$) when hard X-rays were not observed or hard X-ray time profile was complex \citep{1988ApJ...325L..53R,1988ApJ...327..998R}. The authors have pointed out that the correlation between elemental ratios and temperature does not strongly depend on the choice of this time. They have reported the average hard X-ray to 2\,MHz radio peak delay of $\sim$1\,min.

\begin{deluxetable*}{clcLCccccccc}[t]
\tabletypesize{\scriptsize}
\tablecaption{Properties of $^3$He-rich SEP events and associated solar sources.\label{tab:t1}}
\tablehead{\\[-1em]
\colhead{} & \colhead{SEP}& \colhead{Days} & \colhead{$^3$He/$^4$He} & \colhead{Fe/O} & \colhead{Type-III} &
\multicolumn{4}{c}{X-ray Flare} & \colhead{$\log T$} & \colhead{Ref.}\\
\cline{7-10}
\colhead{} & \colhead{Start date}& \colhead{} & \colhead{} &\colhead{} &
\colhead{Start (UT)} & \colhead{Class} & \colhead{Start (UT)} & \colhead{Location} & \colhead{NOAA AR} &\colhead{(K)} &\colhead{}
} 
\startdata 
01& 2010 Jun 12 &  163.75--166.33  & 0.02\pm0.002 &  0.93\pm0.08 & 00:54 &M2.0    & 00:30   & N23W50  & 11081   & 6.65&i\\ 
02& 2010 Sep 02 &  245.92--247.13  & 2.85$\pm$0.76 &  0.78$\pm$0.71 & 10:48 &\nodata    & \nodata   & N24W62  & 11102   & 6.30&i\\ 
03& 2010 Sep 17 &  260.71--261.00 & 0.10$\pm$0.07 &  \nodata & 00:16 &B5.7    & 00:14   & S20W02  & 11108   & 6.32&i\\ 
04& 2010 Oct 17 &  290.88--292.63 & 0.46$\pm$0.04 & 1.27$\pm$0.23& 08:55 &C1.7    & 08:52  & S18W32  & 11112   & 6.44&i, j, k\\ 
05& 2010 Nov 14 &  318.63--321.38 & 0.20$\pm$0.01 & 1.20$\pm$0.18&\phn 23:50$^{\dagger}$ &C1.1 & \phn 23:50$^{\dagger}$ & S23W28 & 11123   & 6.34&i, j, l\\ 
06& 2010 Nov 17 &  321.67--322.42 & 2.75$\pm$0.33 & 1.25$\pm$0.34& 08:08 &B3.4    & 08:07  & S23W72  & 11123   & 6.24&i, j, l\\ 
07& 2011 Jan 27 &  027.88--030.50 & 0.08$\pm$0.01 & 1.15$\pm$0.18 & 08:41 & B6.6  & 08:40  & N14W80  & 11149 & 6.23&j\\ 
08& 2011 Feb 18 &  049.25--051.75 & 0.11$\pm$0.01 & 1.46$\pm$0.10 & \phn 21:32$^{\dagger}$ & C1.1 & \phn 21:30$^{\dagger}$ & S19W46 & 11158 & 6.41&k, m\\ 
09& 2011 Jul 08 & 189.00--190.00 & 0.42$\pm$0.04 & 0.91$\pm$0.15 & \phn 14:27$^{\dagger}$ & B7.6 & \phn 14:25$^{\dagger}$ & N15W91 & 11244 & 6.35&j, n\\
10& 2011 Jul 09 & 190.00--191.00 & 2.35$\pm$0.13 & 1.31$\pm$0.20 & \phn 16:25$^{\dagger}$ & \nodata & \nodata & N16W43 & \nodata & 6.36&n, o\\
11& 2011 Aug 01 & 213.08--213.63 & 0.02$\pm$0.01 & 2.61$\pm$0.47 & \phn 18:58$^{\dagger}$ & C1.7 & \phn 19:01$^{\dagger}$ & N15W50 & 11265 & 6.39&j\\
12& 2011 Aug 26 &  238.46--240.25 & 0.42$\pm$0.03 & 1.51$\pm$0.13 & 00:42 & B4.4  & 00:41  & N15W62  & 11271 & 6.37&j\\ 
13& 2011 Dec 24 &  358.75--359.13 & 0.06$\pm$0.01 & 1.42$\pm$0.26 & 11:55 & C4.9  & 11:25  & N16W92  & 11376 & 6.65&j\\ 
14& 2012 Jan 03 &  003.42--004.08 & 0.16$\pm$0.01 & 1.26$\pm$0.11 & 01:49 & B5.0  & 01:40  & S20W63  & 11386 & 6.32&j\\
15& 2012 May 14 & 135.83--137.13 & 0.02$\pm$0.002 & 0.40$\pm$0.04 & 09:38 & C2.5  & 09:35  & N09W46  & 11476 & 6.56&j, k\\
16& 2012 Jun 08 & 160.71--161.75 & 0.38$\pm$0.04 & 2.00$\pm$0.33 & 07:13 & C4.8  & 07:11  & N13W40 & 11493 & 6.30&j, k\\
17& 2012 Aug 02 & 215.54--217.00 & 0.72$\pm$0.02 & 0.99$\pm$0.09 & 03:00 & B6.4  & 03:00  & N12W58 & 11528 & 6.32&o\\
18& 2012 Nov 20 & 325.46--326.00 & 7.84$\pm$0.75 & 0.92$\pm$0.19 & 01:28 & B6.0  & 01:29  & S18W60 & 11618 & 6.31&l, j, o\\
19& 2013 Jul 17 & 198.13--199.50 & 0.26$\pm$0.02 & 0.84$\pm$0.17 & \phn 20:16$^{\dagger}$ & B1.8 & \phn 20:18$^{\dagger}$ & N19W69 & \nodata & 6.35&j\\
20& 2014 Mar 29 & 088.33--089.00 & 0.11$\pm$0.01 & 0.42$\pm$0.08 & \phn 23:46$^{\dagger}$ & M2.6 & \phn 23:44$^{\dagger}$ & N18W25 & 12017 & 6.57&j\\
21& 2014 Apr 18 & 108.25--108.88 & 0.66$\pm$0.04 & 2.96$\pm$0.24& \phn 21:58$^{\dagger}$ & C3.2 & \phn 21:50$^{\dagger}$ & S15W24 & 12036 & 6.54&j, o\\
22& 2014 May 04 & 124.21--125.75 & 0.11$\pm$0.01 & 0.73$\pm$0.11 & \phn 20:16$^{\dagger}$ & C1.8 & \phn 20:18$^{\dagger}$ & S08W38 & \nodata & 6.33&j\\
23& 2014 May 16 & 136.54--137.75 & 4.69$\pm$0.28 & 2.15$\pm$0.27 & 04:01 & \nodata  & \nodata  & S12W44 & \nodata & 6.31&j, o\\
24& 2015 Feb 06  & 037.25--037.88 & 0.81$\pm$0.10 & 1.80$\pm$0.27 & \phn 23:30$^{\dagger}$ & B6.5 & \phn 23:30$^{\dagger}$ & N14W19 & \nodata & 6.19&o\\
\enddata
\tablecomments{$^3$He/$^4$He, Fe/O from ULEIS/ACE at 0.320--0.453\,MeV\,nucleon$^{-1}$. $^{\dagger}$Previous day from SEP start date.}
\tablerefs{ (i) \citet{2016ApJ...833...63B}; (j) \citet{2015ApJ...806..235N}; (k) \citet{2014SoPh..289.3817R}; (l) \citet{2015AA...580A..16C}; (m) \citet{2018ApJ...869L..21B}; (n) \citet{2014ApJ...786...71B}; (o) \citet{2016ApJ...823..138M}}
\end{deluxetable*}

It is believed that anomalous abundances of $^3$He-rich SEPs are the signature of a unique acceleration mechanism associated with magnetic reconnection in solar flare sites \citep[e.g.,][]{1998SSRv...86...79M}. The models of $^3$He-rich SEPs usually involve temperature in ion acceleration \citep[e.g.,][]{1997ApJ...477..940R,2006ApJ...636..462L,2018ApJ...862....7M,2020ApJ...888...48K}. \citet{1992ApJ...391L.105T} and \citet{1997ApJ...477..940R} have reported a model of $^3$He and heavy-ion acceleration by obliquely propagating electromagnetic ion cyclotron waves. \citet{1997ApJ...477..940R} have found that among discrete values of 2, 4, 6, and 10\,MK, the abundance ratios are close to observed values at temperature 4\,MK. In the model of $^3$He and $^4$He acceleration by parallel-propagating plasma waves, \citet{2006ApJ...636..462L} have found a temperature range of 0.8--3.0\,MK (median and mean 1.4 MK and 1.6\,MK, respectively) by fitting the observed spectra of $^3$He and $^4$He for six events. The mechanism of heavy-ion enhancement in $^3$He-rich events based on charge-to-mass ($Q/M$) dependence of Coulomb energy losses below the Bragg peak ($\sim$1\,MeV\,nucleon$^{-1}$) requires temperatures 0.1--0.6\,MK \citep{2018ApJ...862....7M}. In an approach that combines $Q/M$ dependent Coulomb losses with $Q/M$ dependent stochastic acceleration, the enrichment of heavy and ultra-heavy ions that is similar to the measured one can be obtained with temperatures of $\sim$1\,MK \citep{2008ApJ...681.1653K,2020ApJ...888...48K}. 

The temperature in solar sources of $^3$He-rich SEPs (that can be only determined indirectly) and the effect on ion abundances is largely unexplored. In this paper, we study 24 $^3$He-rich SEP events in the period 2010--2015. To determine the temperature in $^3$He-rich source flares, we use EUV imaging observations that can monitor lower temperatures when ions are not fully stripped. Furthermore, to examine a linkage between the ion abundances and temperature in $^3$He-rich sources, we use low-energy ($\lesssim$0.2\,MeV\,nucleon$^{-1}$) measurements, where ion charge states are dominated by thermal collisions.

\section{Instrumentation and event selection} \label{sec:instr}

The elemental abundances of the $^3$He-rich SEP events in this survey are measured with the time-of-flight mass spectrometer Ultra Low Energy Isotope Spectrometer \citep[ULEIS;][]{1998SSRv...86..409M} on the {\sl Advanced Composition Explorer} \citep[ACE;][]{1998SSRv...86....1S}. Depending on ion species, ULEIS roughly covers the energy range 0.1--10\,MeV\,nucleon$^{-1}$. For example, the Fe measurements can extend down to 0.04\,MeV\,nucleon$^{-1}$. The solar sources of $^3$He-rich SEPs are investigated using high-resolution observations from the Atmospheric Imaging Assembly \citep[AIA;][]{2012SoPh..275...17L} on the {\sl Solar Dynamics Observatory} \citep[SDO;][]{2012SoPh..275....3P}. The AIA provides full-disk images of the Sun with 1.5$\arcsec$ spatial and 12\,s temporal resolution at seven EUV, two UV, and one visible-light band. The AIA data were processed with the standard AIA SolarSoft procedure {\tt aia\_prep.pro} available via SolarSoft library and normalized by the exposure time.

\begin{figure*}
\epsscale{1.17}
\center
\plotone{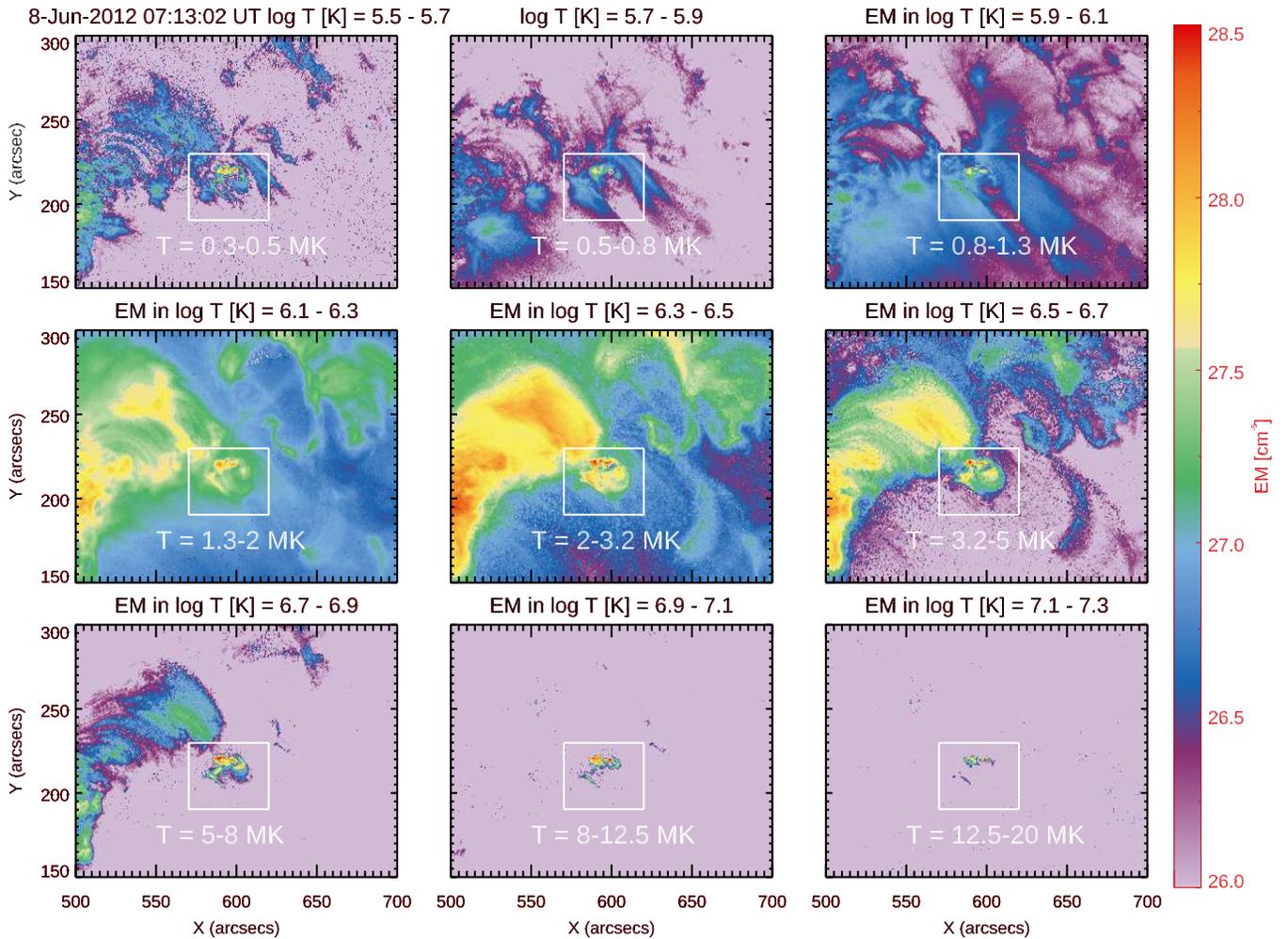}
\caption{Emission Measure (EM) maps for the 2012 June 8 event (event \#16) at the start time of the type-III radio burst. The white rectangle surrounds the source region, where the temperature was determined. Color-coding indicates the total EM contained within a $\log T$ range shown at the top of each panel. \label{fig:f1}}
\end{figure*}

Numerous $^3$He-rich SEP events were measured at L1 during the SDO era in solar cycle 24 \citep{2014SoPh..289.4675R,2014ApJ...786...71B,2016ApJ...833...63B,2018ApJ...869L..21B,2015AA...580A..16C,2015ApJ...806..235N,2016ApJ...823..138M}, however, for this study, we select only some of these events. The reason is that seven previously reported events (2010 Sep 4, Nov 4; 2012 Jan 13; 2013 Dec 24; 2014 Feb 6, Apr 24, May 29) have the source close or behind the limb with weak or no activity observed from the Earth. Another five events (2010 Oct 19; 2011 Dec 14; 2012 Nov 18, Jul 3; 2014 Jan 1) have the AIA images saturated, and not usable for further analysis. Table~\ref{tab:t1} lists the properties of the examined 24 $^3$He-rich SEP events and the associated solar sources. Column 1 indicates the event number, column 2 the SEP event start date, and column 3 the time intervals used for computing elemental ratios based on the ion intensities near 0.2\,MeV\,nucleon$^{-1}$. Columns 4 and 5 provide the event integrated (averaged) $^3$He/$^4$He and Fe/O ratios, respectively, at the energy range of 0.320--0.453\,MeV\,nucleon$^{-1}$ where ion abundances have been often reported. Column 6 gives the type-III radio burst start time from WAVES/Wind \citep{1995SSRv...71..231B} data with 60\,s cadence. Columns 7 and 8 indicate the GOES 14--15 \citep{10.1117/12.254076} X-ray flare class and start time, respectively. Column 9 gives the flare site in AIA images. Column 10 shows the NOAA active region (AR) number. The X-ray flare class, start time, and NOAA AR number were taken from the Edited Solar Events, compiled by the NOAA Space Weather Production Center (\url{ftp://ftp.swpc.noaa.gov/pub/warehouse}). Column 11 provides a decadic logarithm of the temperature in the solar source, derived from AIA channels by the differential emission measure (DEM) analysis. Specifically, the column gives the EM-weighted temperature (see Section~\ref{subsec:td}). The last column indicates references to earlier studies of these events.

\begin{figure*}
\epsscale{1.15}
\plotone{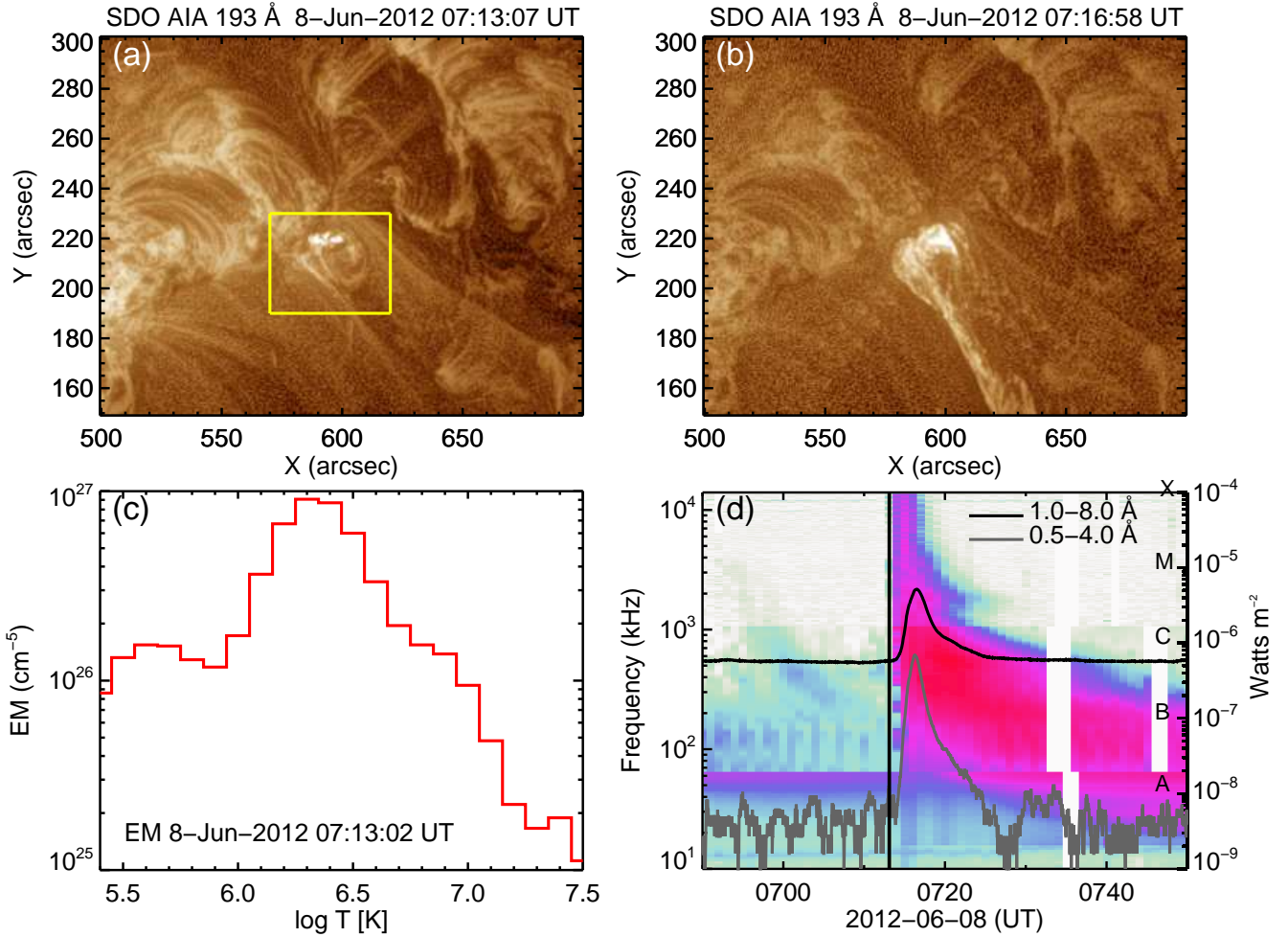}
\caption{ (a) AIA/SDO 193\,{\AA} image for event \#16 with the solar source in the yellow box at the start time of the type-III radio burst. (b) AIA/SDO 193\,{\AA} image for event \#16 4\,min later with a jet in the solar source. To enhance small-scale structures, the images are processed with the Multiscale Gaussian Normalization code {\tt mgn.pro} \citep{2014SoPh..289.2945M}. (c) The average EM profile for the solar source in the yellow box at the start time of the type-III radio burst. (d) WAVES/Wind radio spectrogram and GOES-15 X-ray fluxes (two curves). Color-coding in the radio spectrogram represents $E-$intensity (dB $>$ background). The white patches in the radio spectrum mean the data gap. The vertical line marks the time of the temperature determination. The labels A, B, C, M, and X mark flare classes in the 1--8\,{\AA} channel. \label{fig:f2}}
\end{figure*}

In most of the investigated events (17 out of 24), there was an EUV jet observed in the associated solar source. Another type of activity includes EUV coronal wave (events \#1, \#4, \#11), ejection (events \#13, \#21) or just EUV brightening (events \#5, \#22). Except for events \#17 and \#24, the solar sources for the remaining events have been identified in the earlier papers.

\section{Results} \label{sec:res}

\subsection{Temperature determination} \label{subsec:td}

In this work, we determine the temperature at the start time of type-III radio bursts as observed by WAVES/Wind at 14 MHz that corresponds to $\sim$2\,R$_\sun$ according to the electron density model \citep{1999A&A...348..614M}. The average temperature is extracted from a rectangular region around the source that shows EUV brightening at the type-III onset. In cases when the jet is the source, the region covers both the jet foot-point and the jet spire. Note that the EUV jet was not always well visible at the time of type-III radio burst, and the only activity observed was EUV brightening (events \#14, 15, 16, 19, 20, 23). See Figure~\ref{fig:f2} for event \#16.

We perform a DEM analysis to study the thermal structure of the plasma in the solar sources of $^3$He-rich SEPs. The DEM is a physical quantity related to the electron density and the temperature gradient of a plasma. It is determined from observations of emission lines on the assumptions that plasma is optically thin and in ionization equilibrium. The DEM is reconstructed from the multi-wavelength observations and the instrumental temperature response by the inversion method. We use the DEM inversion developed by \citet{2015ApJ...807..143C} for EUV AIA imaging observations. We coalign AIA images in six channels (94, 131, 171, 193, 211, and 335\,{\AA}), which are sensitive to a range of coronal temperatures and use them as input in the DEM analysis. The method computes the temperature responses for six AIA channels (with the SolarSoft function {\tt aia\_get\_response.pro}), using the CHIANTI atomic database \citep{2019ApJS..241...22D} and coronal abundances \citep{1992PhyS...46..202F,1998SSRv...85..161G,2002ApJS..139..281L}. The DEM analysis is performed in the solar source and its surrounding area. The EM (the integral of DEM) is computed for each pixel position in the temperature interval $\log T$\,[K] $=$ 5.4--7.5 (0.3--30\,MK). Finally, the EM is spatially averaged over the rectangular region around the source. 

Figure~\ref{fig:f1} illustrates the DEM analysis for the $^3$He-rich event 2012 June 8 (event \#16). The figure shows EM maps at various temperature intervals. The solar source is surrounded by a white box. The box is positioned to cover both the hot core (see map at 3.2--5\,MK) as well as two cooler peripheral structures (see map 0.5--0.8\,MK) adjacent to the core that resembles open field lines. Note that an EUV jet was later ejected along one of those elongated structures (see Figure~\ref{fig:f2}). Event \#16 has one of the highest $^3$He/$^4$He and Fe/O ratios in our sample (Table~\ref{tab:t1}). Only events \#21 and \#23 show the higher common $^3$He and Fe enrichment. 

Figure~\ref{fig:f2} displays 193\,{\AA} images of the solar source for event \#16 at the start time of the type-III radio burst (panel a) and a later time (panel b) when a jet was clearly seen. Figure~\ref{fig:f2}(c) plots the average EM curve calculated for the yellow boxed region shown in Figure~\ref{fig:f2}(a). The average EM curve peaks at $\log T$\,[K] = 6.3 ($\sim$2\,MK) with peak EM = 9.12$\times$10$^{26}$\,cm$^{-5}$. There is also low temperature plasma available at some pixel locations at the source
region. Figure~\ref{fig:f2}(d) shows the WAVES/Wind radio spectrogram and the GOES X-ray fluxes. The WAVES data are read and displayed using the package {\tt xwavesds.pro} available at \url{https://solar-radio.gsfc.nasa.gov/data/wind/idl/}. At the highest WAVES frequencies, the type-III radio burst onset is observed at 07:13\,UT and the emission peaked at 07:15\,UT. According to the Edited Solar Events, the type-III radio burst with the start time at 07:13\,UT was observed by Learmonth Solar Observatory at a frequency range of 25--180\,MHz. Using data in Figure~\ref{fig:f2}(c), we calculate the EM-weighted average temperature \citep[see Eqn. 15 in][]{2015ApJ...807..143C} and find that it is $\log T$\,[K] = 6.3. The DEM inversion code gives temperature within an error of 20\%. The errors arise from a number of instrumental effects and photon counting statistics. 

\begin{figure}
\epsscale{1.15}
\plotone{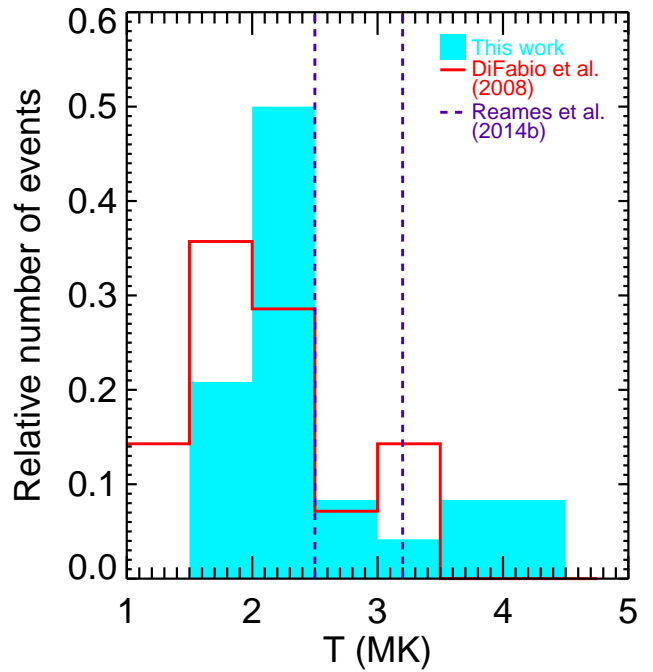}
\caption{Histogram of solar source temperatures for the $^3$He-rich SEP events in this work (cyan shaded) and from \citet{2008ApJ...687..623D} (red line). Two vertical lines indicate the temperature range reported by \citet{2014SoPh..289.4675R}. \label{fig:f3}}
\end{figure}

\begin{figure*}
\epsscale{1.14}
\plotone{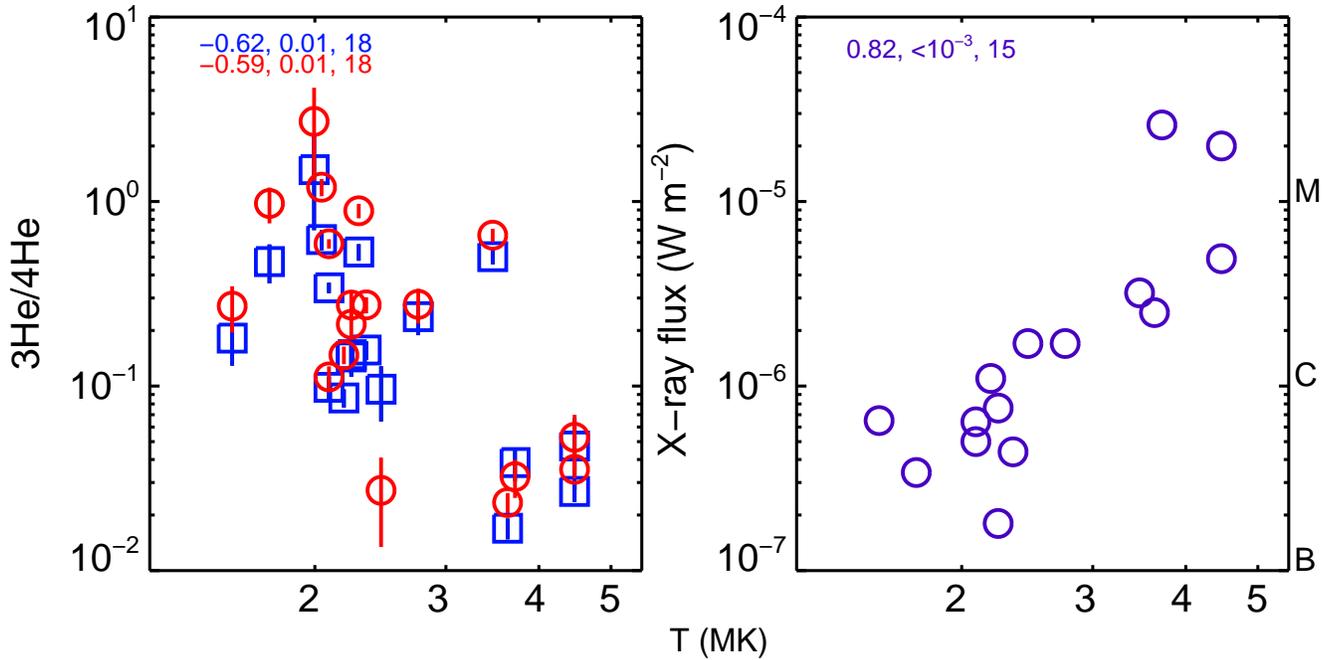}
\caption{The event integrated 0.113--0.160\,MeV\,nucleon$^{-1}$ (blue squares) and 0.160--0.226\,MeV\,nucleon$^{-1}$ (red circles) $^3$He/$^4$He ratio from ULEIS/ACE ({\sl left panel}) and corresponding GOES 1--8\,{\AA} X-ray peak flux ({\sl right panel}) vs. the EM-weighted temperature $T$ in the solar source from AIA/SDO. A triple of numbers means the Spearman rank correlation coefficient, the two-sided significance of its deviation from zero, and the number of events. The letters B, C, M, mark the GOES X-ray flare class. Note, the $^3$He/$^4$He $\sim$4$\times$10$^{-4}$ in the solar wind \citep{1998SSRv...84..275G} for comparison with $^3$He/$^4$He ratios in the left panel. \label{fig:f4}}
\end{figure*}

Examining 24 events in this study, we obtain a temperature range of 1.5--4.5\,MK with an average value of 2.5\,MK and a median of 2.2\,MK. A multi-thermal plasma is found in the solar source for event \#20 where the EM curve shows a double peak with nearly the same intensities at $\log T$\,[K] = 6.4 (2.5\,MK) and $\log T$\,[K] = 6.9 (7.9\,MK). Figure~\ref{fig:f3} plots the histogram of the temperature distribution for the investigated $^3$He-rich events. The histogram shows that the temperature distribution peaks at 2.0--2.5\,MK. The majority of $^3$He-rich SEP events (17 out of 24) have source temperature below 2.5\,MK. These events are associated with small X-ray flares (mostly B-class; see Figure~\ref{fig:f4} or Table~\ref{tab:t1}). Also added is the histogram of inferred temperatures from Fe charge state measurements reported by \citet{2008ApJ...687..623D} as well as the temperature range 2.5--3.2\,MK determined from the abundance enhancement \citep{2014SoPh..289.4675R}. Note that four events of our study with source temperatures 2.0, 2.6, 2.8, and 3.6\,MK are in the survey of 111 Fe-rich impulsive events of \citet{2014SoPh..289.3817R}. The asymmetric shape of the two histograms may suggest a lower temperature boundary in $^3$He-rich solar sources.

\subsection{$^3$He abundance versus temperature} \label{subsec:3heT}

To examine the relationship between the elemental ratios and the temperature, we use only those events from Table~\ref{tab:t1} that are not affected by the background from the preceding event. Especially at low energies, small $^3$He-rich event ion intensities can be contaminated by intensity enhancement from prior large SEP, corotating, or IP shock events. \citet{2004ApJ...606..555M} have discussed that $^4$He is the most susceptible element for contamination, Fe and O are less susceptible, and $^3$He is the least susceptible. In events \#3, \#7, \#8, \#18, $^4$He shows no obvious increase from prior enhancement, and therefore $^3$He/$^4$He presents a lower limit (real $^3$He enrichment is higher). In events \#16, \#22, $^3$He remained enhanced from the previous activity, and the calculated ratio is only an upper limit. In events \#8, and \#20, O shows no increase, and thus Fe/O presents a lower limit. Furthermore, only the ratios with a relative error $\le$0.75 are used.

\begin{figure*}
\epsscale{1.12}
\plotone{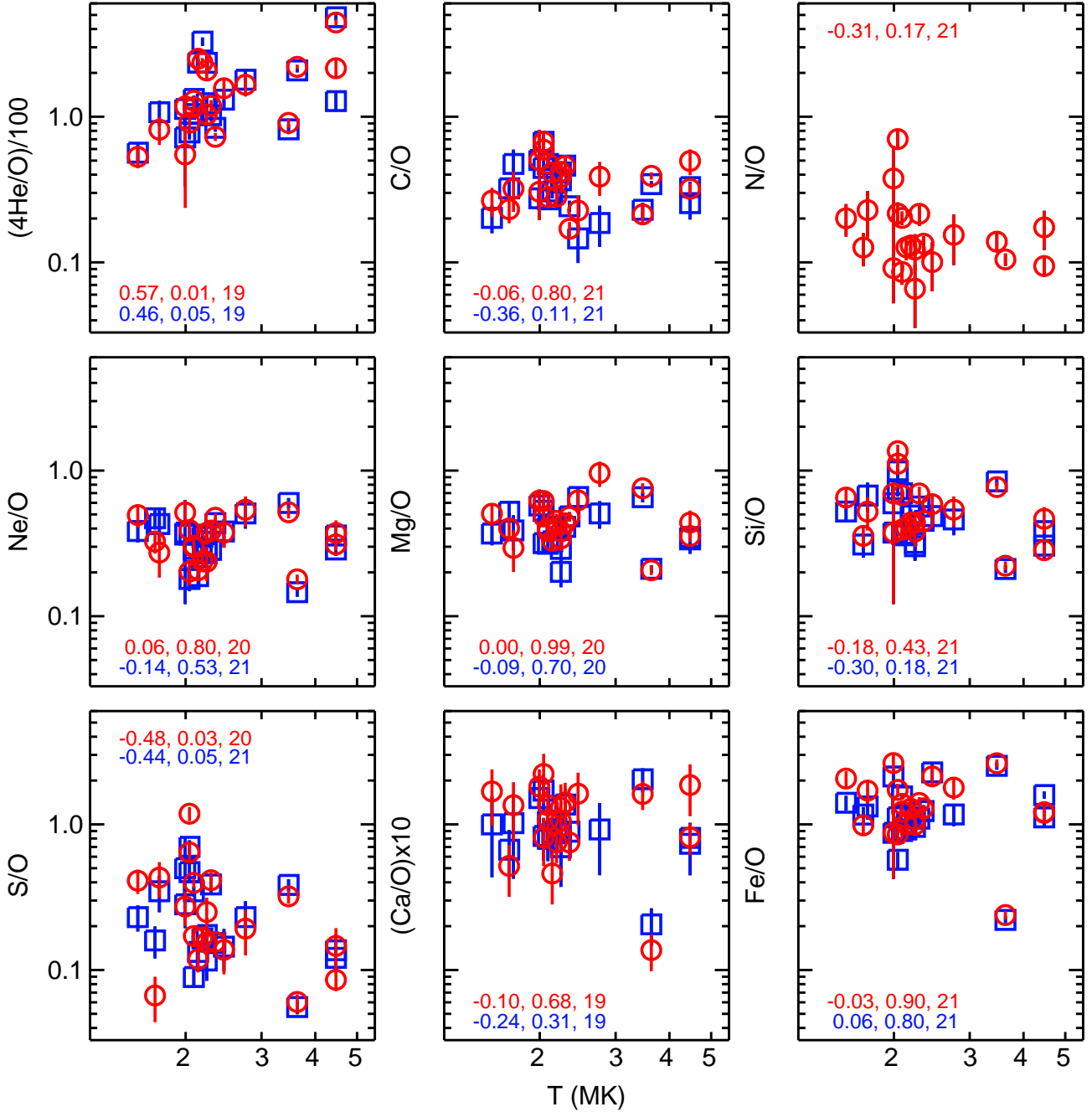}
\caption{The event integrated 0.113--0.160\,MeV\,nucleon$^{-1}$ (blue squares) and 0.160--0.226\,MeV\,nucleon$^{-1}$ (red circles) elemental ratios for different ion species from ULEIS/ACE vs. the EM-weighted temperature $T$ in the solar source from AIA/SDO. A triple of three numbers has the same meaning as in Figure~\ref{fig:f4}. Since N ion is not well resolved at 0.113--0.160\,MeV\,nucleon$^{-1}$, only the high energy ratio is shown. \label{fig:f5}}
\end{figure*}

Figure~\ref{fig:f4}(left) shows the event integrated $^3$He/$^4$He ratio in the two lowest energy channels, 0.113--0.160\,MeV\,nucleon$^{-1}$ and 0.160--0.226\,MeV\,nucleon$^{-1}$, measured by ULEIS versus the EM-weighted temperature $T$ in the $^3$He-rich SEP sources obtained from AIA observations. There is a relatively high anti-correlation between these two quantities, with Spearman rank correlation coefficient of $-$0.62 (0.113--0.160\,MeV\,nucleon$^{-1}$) or $-$0.59 (0.160--0.226\,MeV\,nucleon$^{-1}$) with a low probability (p$=$0.01) of resulting from the random population. The obtained relation indicates a decrease of $^3$He enrichment with a source temperature. This could not be caused by charge-changing processes as the He isotopes would be fully ionized at the inferred temperatures (see Figure~\ref{fig:f7}(left) for $^4$He). They would be fully ionized even at $\sim$0.1\,MK. To explore this result further, we plot in Figure~\ref{fig:f4}(right) GOES 1--8\,{\AA} X-ray peak flux versus the same temperature $T$ as for events in Figure~\ref{fig:f4}(left). The three events are not shown because they have an X-ray flux below the background level. Note that X-ray flux refers to the whole solar disk while the temperature just to the small region around the source. Data in Figure~\ref{fig:f4}(right) indicate a strong positive correlation between X-ray peak flux and the temperature $T$ in the examined $^3$He-rich SEP sources. Thus, a negative trend between the $^3$He/$^4$He and the temperature may represent a dependence on X-ray (EUV) peak flux, wherein large flares produce small $^3$He enrichment. Previously, \citet{1988ApJ...327..998R} have reported an anti-correlation between 1.3--1.6\,MeV\,nucleon$^{-1}$ $^3$He/$^4$He and soft X-ray peak flux with correlation coefficients $-$0.66 (ISEE) and $-$0.55 (GOES) that are surprisingly similar to our values. Recall, that the authors find no evidence for a correlation between $^3$He/$^4$He and soft X-ray temperature (10--16\,MK) at the time of hard X-ray maximum.

\begin{figure*}
\epsscale{1.12}
\plotone{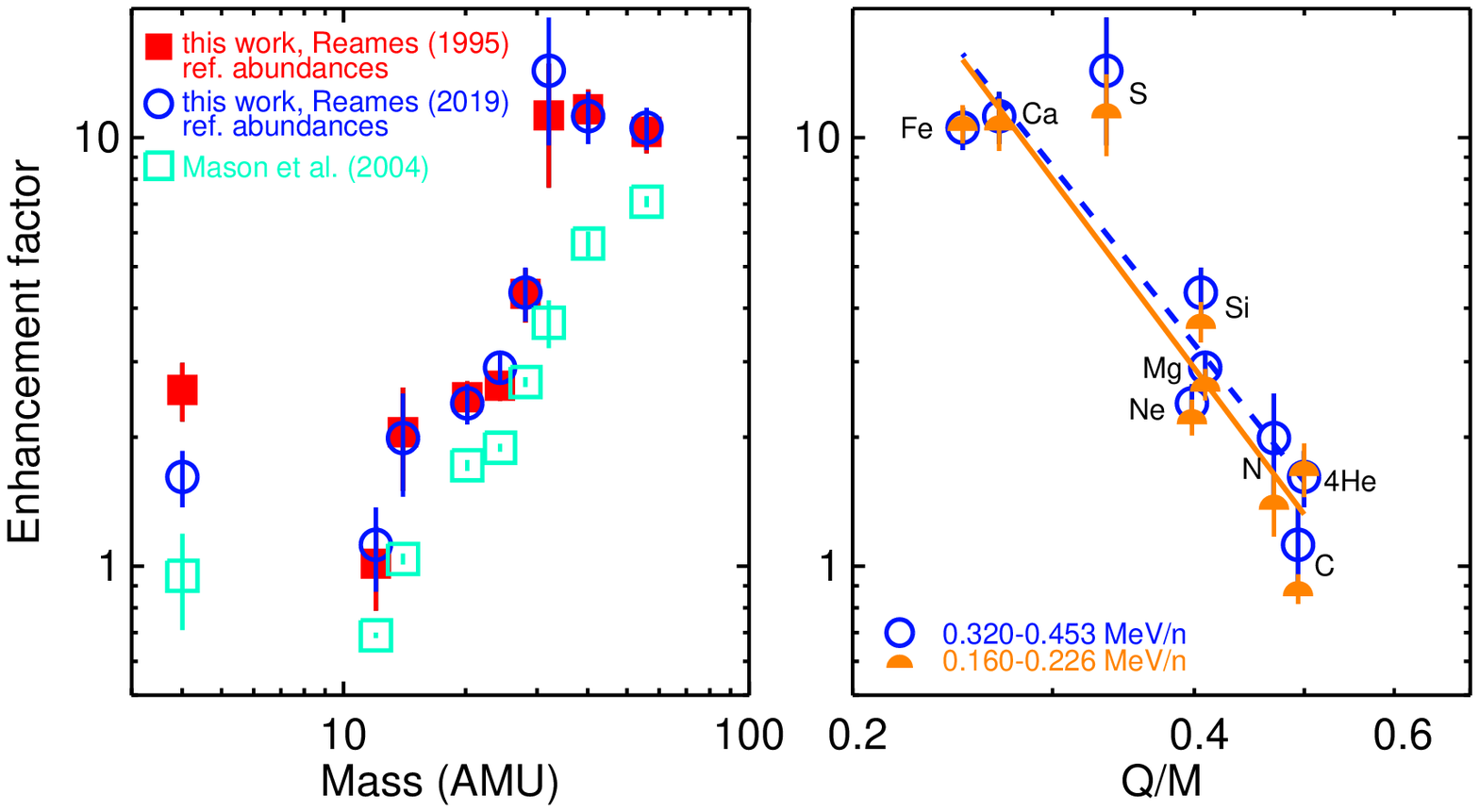}
\caption{{\sl Left panel}: Abundance enhancement relative to O for $^3$He-rich SEP events in this survey at 0.320--0.453\,MeV\,nucleon$^{-1}$, relative to coronal abundances from \citet{1995AdSpR..15g..41R} (red, filled squares) and \citet{2019Atoms...7..104R} (blue circles). Green open squares are values from \citet{2004ApJ...606..555M}. {\sl Right panel}: Abundance enhancement relative to O for $^3$He-rich SEP events in this survey at 0.160--0.226\,MeV\,nucleon$^{-1}$ (orange, filled half-circles) and 0.320--0.453\,MeV\,nucleon$^{-1}$ (blue circles), relative to coronal abundances \citep{2019Atoms...7..104R} vs. $Q/M$ ratio, using charge states calculated for the equilibrium plasma at 2.2\,MK. The solid (dashed) line is a power-law fit that has a slope of $\gamma=-$3.53 ($-$3.31). \label{fig:f6}}
\end{figure*}

\subsection{Heavy-ion abundances versus temperature} \label{subsec:heaT}

Figure~\ref{fig:f5} explores the relation between the event integrated heavy-ion abundances relative to oxygen and the temperature in the solar source for two energy channels 0.113--0.160 \,MeV\,nucleon$^{-1}$ and 0.160--0.226\,MeV\,nucleon$^{-1}$. Only two ratios show a clear relation to the temperature in the $^3$He-rich SEP source, $^4$He/O and S/O. The $^4$He/O shows a relatively high positive (0.46 and 0.57) and S/O a moderate negative ($-$0.44 and $-$0.48) correlation with the temperature. Both these correlations have high statistical significance (p$\le$0.05). There is a weak negative correlation for N/O and Si/O, albeit with a low significance. The C/O is inconclusive, showing no correlation for the higher energy channel and a weak negative trend for the lower energy channel. All other ratios (Ne/O, Mg/O, Ca/O, Fe/O) show no correlation with temperature.

\citet{2008ApJ...687..623D} have reported a positive, moderate correlation (with a coefficient 0.46) between 0.24\,MeV\,nucleon$^{-1}$ Fe/O ratio and 0.062--0.110\,MeV\,nucleon$^{-1}$ Fe charge states ($Q\sim$11--16) for a sample of 14 $^3$He-rich SEP events. The authors pointed out that it is puzzling why the Fe/O ratio becomes larger towards the fully-stripped conditions ($Q/M\sim$0.5). \citet{2000AIPC..528..131M} have also found an increase of Fe/O (0.58--2.3\,MeV\,nucleon$^{-1}$) ratio with Fe charge state (0.18--0.43\,MeV\,nucleon$^{-1}$) for seven $^3$He-rich SEP events. The authors conclude that such correlation currently has no satisfactory explanation. The reported correlations are inconsistent with our results. If we replace (not shown) in Figure~\ref{fig:f5} the temperature with the computed Fe charge states ($Q\sim$11--16 for the range of observed temperatures), the correlation between the Fe/O ratio and Fe charge state remains zero. 

\subsection{Abundance enhancement factor} \label{subsec:ef}

Figure~\ref{fig:f6}(left) plots the abundance enhancement factor (X/O)/(X/O)$_{\mathrm{coronal}}$ of element X for the $^3$He-rich SEP events at 0.320--0.453\,MeV\,nucleon$^{-1}$ versus mass. The enhancement factor is calculated from the unweighted average abundances divided by the mean coronal abundances taken from \citet{1995AdSpR..15g..41R} and \citet{2019Atoms...7..104R} as measured by 5--12\,MeV\,nucleon$^{-1}$ gradual SEP events. All abundances are normalized to oxygen. For comparison, shown are the values from \citet{2004ApJ...606..555M} survey of 20 $^3$He-rich SEP events that are in the same energy bin and the reference coronal abundances from \citet{1995AdSpR..15g..41R}. 

\begin{figure*}
\epsscale{1.13}
\plotone{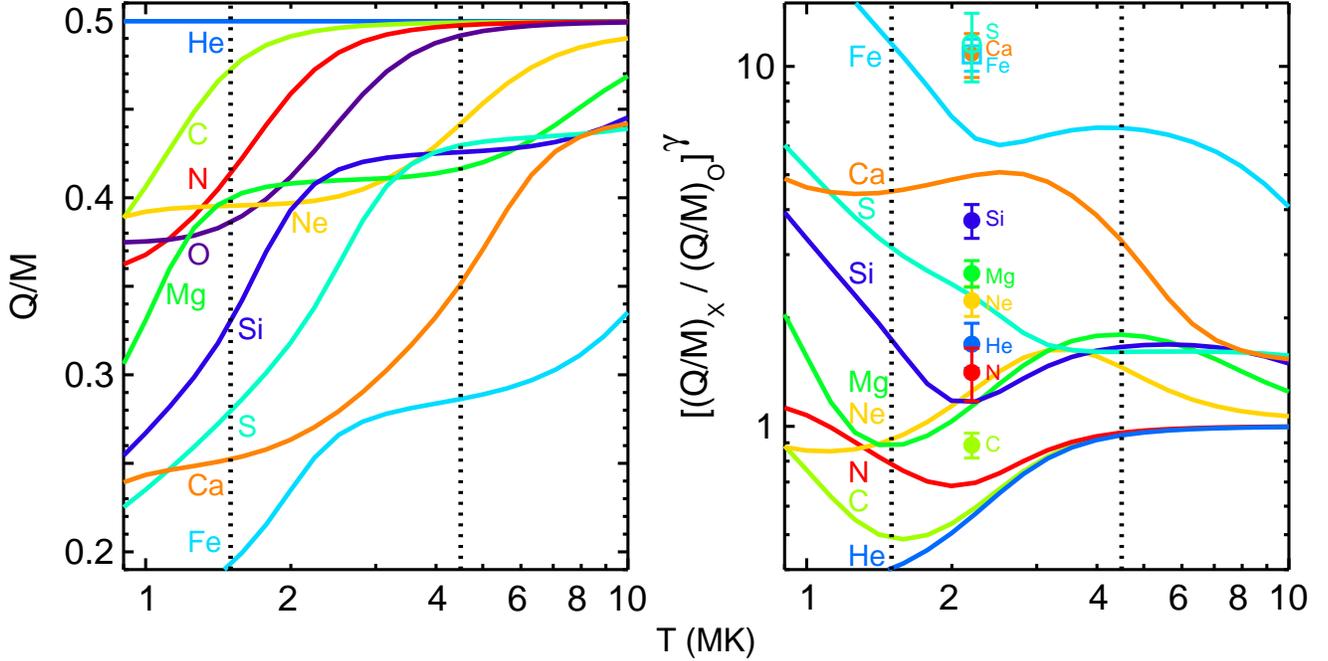}
\caption{A $Q/M$ ratio ({\sl left panel}) and an enhancement factor at 0.160--0.226\,MeV\,nucleon$^{-1}$ ($\gamma=-$3.53) for a selected element X ({\sl right panel}) vs. temperature $T$ as calculated at ionization equilibrium. Measured enhancements (filled circles) are plotted at 2.2\,MK. The two vertical dotted lines mark the range of temperatures obtained for the examined $^3$He-rich sources. \label{fig:f7}}
\end{figure*}

A noticeable feature is that the survey presented here shows a higher enhancement factor for all elements with the highest deviation for $^4$He and $^{32}$S. Furthermore, the average (0.95$\pm$0.31) and median (0.42) values of $^3$He/$^4$He ratio at 0.320--0.453\,MeV\,nucleon$^{-1}$ are more than an order of magnitude higher than values reported by \citet{2004ApJ...606..555M}. A probable reason for such discrepancies is in the selection of the events. \citet{2004ApJ...606..555M} have focused on $^3$He-rich flares with measurable ultra-heavy ions (78--220\,AMU). It raises the question of whether $^3$He-rich SEPs with ultra-heavy populations present a specific subclass of the events. Contrary to \citet{2004ApJ...606..555M}, $^{32}$S, $^{40}$Ca, $^{56}$Fe, are not ordered by the mass but show similar abundance enhancement. The same has been reported by \citet{1995AdSpR..15g..41R} for Ca, Fe. Note that the $Q/M$ dependence could be much more complicated than a monotonically changing function \citep{2001ApJ...563..403D}. \citet{2016ApJ...823..138M} have reported 16 $^3$He-rich SEP events with an enhanced S/O ratio from the period 1999--2015. This study includes six of these events (six out of seven in the solar cycle 24), while the \citet{2004ApJ...606..555M} study involves only one such event. It is an obvious reason for the difference in S/O enhancement between the two surveys. The high values of $^4$He/O ratio (from $\sim$133 to $\sim$433 at 0.320--0.453\,MeV\,nucleon$^{-1}$) have been reported previously in four $^3$He-rich SEP events from nearly quiet Sun in 2007--2008 \citep{2009ApJ...700L..56M}. The average $^4$He/O of this study in the same energy range is 147$\pm$22 which is a factor of 2.7 higher than the value reported by \citet{2004ApJ...606..555M}.

Figure~\ref{fig:f6}(right) plots the abundance enhancement factor at energy ranges 0.160--0.226\,MeV\,nucleon$^{-1}$ and 0.320--0.453\,MeV\,nucleon$^{-1}$, using the reference coronal abundances from the latest compilation reported by \citet{2019Atoms...7..104R}, versus $Q/M$ ratio, using charge states calculated for the equilibrium plasma at 2.2\,MK, the median temperature for the examined $^3$He-rich SEP sources.  To compute the charge states, we use the CHIANTI ionization equilibrium tables that are a compilation of the ionization fractions from different works \citep[e.g.,][]{1985A&AS...60..425A,1992ApJ...398..394A,1998A&AS..133..403M}. A power-law ($Q/M$)$^{\gamma}$ fit to data points is shown and has a slope $\gamma=-$3.53 for 0.160--0.226\,MeV\,nucleon$^{-1}$ and $\gamma=-$3.31 for 0.320--0.453\,MeV\,nucleon$^{-1}$. The value of  $\gamma$ at 0.320--0.453\,MeV\,nucleon$^{-1}$ is in excellent agreement with the slope ($\gamma=-$3.26) reported earlier by \citet{2004ApJ...606..555M}, where $Q/M$ ratios are calculated for 3.2\,MK equilibrium plasma and the fit includes ultra-heavy species. A similar value of a power-law index $\gamma=-$3.64 at 3--10\,MeV\,nucleon$^{-1}$ has been reported by \citet{2014SoPh..289.3817R}, exploring 111 Fe-rich impulsive events.

To understand the variations of abundance ratios with temperature in Figure~\ref{fig:f5}, we examine the enhancement factor \citep[e.g.,][]{1999GeoRL..26..149C} as a function of temperature \citep{2004ApJ...606..555M}. It can be obtained from Figure~\ref{fig:f6}(right) that the enhancement factor for element X is [($Q/M$)$_{\mathrm{X}}$/($Q/M$)$_{\mathrm{O}}$]$^{\gamma}$ with O as a baseline element. Figure~\ref{fig:f7}(left) shows the calculated equilibrium $Q/M$ ratios as a function of temperature for selected elements using the CHIANTI atomic database. Note that for a particular element, the $Q$ is the mean charge state computed using the probabilities of all possible charge states at a given temperature. The $Q/M$ ratio profiles in Figure~\ref{fig:f7}(left) were used to calculate the enhancement factor relative to O. Figure~\ref{fig:f7}(right) plots the calculated enhancement factor over a range of temperatures for 0.160--0.226\,MeV\,nucleon$^{-1}$ using power-law index $\gamma=-$3.53. The measured enhancements in the energy range 0.160--0.226\,MeV\,nucleon$^{-1}$ are plotted at 2.2\,MK. The largest deviation between predicted and measured enhancements are for S, Si, and $^4$He with factors of 5.0, 3.2, and 3.0, respectively. All other elements (C, N, Ne, Mg, Ca, Fe) show reasonable agreement with the deviations ranging between 1.5 to 2.3. Note that over the inferred temperature range also $^4$He and Si show reasonable agreement. \citet{2004ApJ...606..555M} have reported a better agreement where the largest deviations have a factor of 1.7. 

Figure~\ref{fig:f7}(right) demonstrates that the calculated enhancement factor for Fe, Ca, Si, Ne, and N has a complex profile with both decreasing and increasing parts in the temperature range of this survey. For example, Fe may have the same enhancements at 2.1\,MK and 4.5\,MK; or Si at 1.5\,MK and 4.5\,MK. It suggests that no simple trend may exist between the abundances of these elements (when normalized to oxygen) and the temperature. Indeed, the measurements in Figure~\ref{fig:f5} reveal no correlation for Fe/O, Ca/O, Ne/O, or the correlation is not statistically significant for N/O, Si/O. The statistically significant, moderate anti-correlation for S/O appears to be consistent with the monotonously decreasing profile of the enhancement factor computed for S. Note that the survey averaged enhancement for S shows the largest deviation from the predicted value. This is likely a consequence of the extreme S-enrichment which might not follow a simple ($Q/M$)$^{\gamma}$-dependent pattern (see Figure~\ref{fig:f6}(right)). The calculated enhancement factor for He, C, and Mg shows a relatively simple increase in the measured temperature range, predicting a positive trend with temperature. For He, the calculated enhancement factor equals the ratio with a constant numerator ($Q/M$)$_{\mathrm{He}}$ and increasing denominator ($Q/M$)$_{\mathrm{O}}$ (see Figure~\ref{fig:f7}(left)) where the ratio has a negative exponent $\gamma$. This leads to an increase in calculated He enhancement shown in Figure~\ref{fig:f7}(right). The measurements in Figure~\ref{fig:f5} show that only $^4$He/O has a significant positive correlation with temperature, C/O, and Mg/O do not. 

\section{Discussion} \label{sec:di}

The remote-sensing approach with EUV observations performed for $^3$He-rich SEP sources at the time of type-III radio bursts provides a temperature range of 1.4--4.5\,MK with a median of 2.2\,MK. This value is in good agreement with a median temperature of 2.1\,MK derived from in-situ Fe ion charge states at 0.062--0.110\,MeV\,nucleon$^{-1}$ \citep{2008ApJ...687..623D}. This agreement, on the other hand, implies that $^3$He-rich SEPs in this survey may be accelerated very close to the type-III radio burst onset times. Our temperature range is also consistent with the temperature of the source plasma (2.5--3.2\,MK) that is accelerated in $^3$He-rich SEP events as deduced from the abundance measurements in 3--5\,MeV\,nucleon$^{-1}$ \citep{2014SoPh..289.3817R,2015SoPh..290.1761R}.

We found a significant negative correlation between the $^3$He/$^4$He ratio and the temperature in $^3$He-rich sources that may be associated with the strong link between the temperature and X-ray (EUV) peak flux. It is consistent with earlier work that has shown that 1.3--1.6\,MeV\,nucleon$^{-1}$ $^3$He/$^4$He \citep{1988ApJ...327..998R} as well as 1--10\,MeV\,nucleon$^{-1}$ ultra-heavy ion enhancement \citep{2004ApJ...610..510R} anti-correlates with soft X-ray peak intensity. The authors argued that in small flares, much of the energy is absorbed by the rare $^3$He and ultra-heavy elements, while in large flares, there is enough energy for the acceleration of more abundant elements, decreasing the relative abundances of the rare species. Remarkably, the negative trend between $^3$He/$^4$He and the temperature found here does not show up at higher temperatures ($>$10\,MK) as reported by \citet{1988ApJ...327..998R}. \citet{2006ApJ...636..462L} have reported a decrease of the $^3$He/$^4$He ratio with increasing background plasma temperature as a result of thermal damping of waves that accelerate $^4$He. In their model \citep[Figure~9, left in][]{2006ApJ...636..462L}, the $^3$He/$^4$He ratio decrease slows down at higher temperatures and almost stops at $>$10\,MK. Thus, the negative correlation at low temperatures found in this study and zero correlation at high temperatures reported by \citet{1988ApJ...327..998R} may well correspond with this model.

The abundances of heavier elements, relative to oxygen, generally behave with the temperature as expected from the calculated enhancement factor using equilibrium charge states. The two elements, C and Mg, are an exception. Their relative abundances show no relation with temperature, although the computed enhancement factor suggests a positive trend. The C and Mg, together with $^4$He, are the elements with the smallest or no change in $Q/M$ in the measured temperature range (see Figure~\ref{fig:f7}(left)). However, this probably does not explain the missing temperature dependence for the relative abundances of C and Mg as $^4$He shows the strongest one. We may speculate that the processes (other than charge-changing) that cause a negative correlation for the $^3$He/$^4$He ratio might eliminate the presumed positive correlation for C/O and Mg/O. Specifically, the effect of the size of the flare on less abundant C and Mg in comparison to O would cause the C/O and Mg/O to decrease with the temperature. The same effect would cause $^4$He/O to increase with temperature.

The choice of the time and location for the temperature determination may introduce uncertainty to the relationship between the temperature and ion abundances. It is unknown at what time ions are accelerated in $^3$He-rich SEP events. It is believed that it is in the impulsive phase of solar flares along with non-relativistic ($\sim$10--100\,keV) electrons \citep{1985ApJ...292..716R,2012ApJ...759...69W}. Timing analysis of these small events would not be helpful due to significant errors in estimations of ion solar release time that can be tens of minutes \citep{2000ApJ...545L.157M,2016A&A...585A.119W}. Furthermore, energetic ions in $^3$He-rich SEP events do not show optical signatures as energetic electrons that could narrow down the time of the ion acceleration. Also, it is not well-known at what particular place at the flare sites ions are accelerated (jet spire or foot-points). If particles are accelerated at the cooler jet spire, it is probably a larger area, involving the hotter foot-point region, that affects the ion charge states. Some recent works have investigated the temperature of jets associated with type-III radio bursts \citep{2013ApJ...769...96C,2016A&A...589A..79M,2019A&A...632A.108M}. \citet{2016A&A...589A..79M} have examined 20 AR jets and reported the temperature of the jet spire in the range of 1.6--2.0\,MK. In a study of one AR jet, \citet{2019A&A...632A.108M} found a similar low temperature of 2\,MK at the spire and 5--8 MK hot plasma at the foot-point region. \citet{2013ApJ...769...96C} have examined the temporal evolution of the temperature in one jet. The authors have reported 7\,MK hot foot-points and cooler (1.3--2.5\,MK) jet spire at type III-radio burst onset. The temperature distribution in this survey peaks at slightly higher temperatures (2.0--2.5\,MK) compared to the temperatures in the jet spire reported by \citet{2013ApJ...769...96C} and \citet{2016A&A...589A..79M}.

We confirm steep power-law dependence of abundance enhancement factor on $Q/M$ for heavy-ion species ($^4$He--Fe) as previously reported by \citet{2004ApJ...606..555M} and \citet{2014SoPh..289.3817R}. The slope of the power-law dependence is in good agreement with these studies that involve ultra-heavy elements. The trend seen in the abundance enhancement factor is consistent with the most recent models on heavy ion acceleration by cyclotron resonance, showing that the efficiency of acceleration increases inversely with the $Q/M$ \citep{2017ApJ...835..295K,2020ApJ...890..161F}. In the model of ion scattering on reconnecting magnetic islands \citep{2009ApJ...700L..16D}, the heavy-ion production rate can be expressed as a power-law in $Q/M$.

\section{Conclusions} \label{sec:co}

We have examined 24 $^3$He-rich SEP events and for the first time (i) determined temperature in the event associated source flare using EUV imaging observations and (ii) analyzed the relationship between the suprathermal ion abundances and the EUV temperature. A similar analysis for $>$1\,MeV\,nucleon$^{-1}$ ions was performed more than 30 years ago with temperature inferred from the soft X-ray observations of the whole solar disk. The study presented here shows
\begin{enumerate}
	\item a remarkably good agreement between temperature inferred from EUV images and the temperature obtained from ion charge states and abundance measurements,
	\item a negative correlation between $^3$He/$^4$He ratio and the source temperature that is apparently caused by non-charge changing mechanisms,
	\item the behavior of the elemental ratios $^4$He/O, N/O, Ne/O, Si/O, S/O, Ca/O, Fe/O with the source temperature as expected from the calculated enhancement factor assuming ionization equilibrium.
\end{enumerate}

\acknowledgments

Work at SwRI was supported by NASA grants 80NSSC19K0079 and 80NSSC20K1255, and at JHU/APL by NASA grant NNX17AC05G/125225. This study benefits from discussions within the International Space Science Institute (ISSI) Team ID 425 ‘Origins of $^3$He-rich SEPs’. CHIANTI is a collaborative project involving George Mason University, the University of Michigan (USA), University of Cambridge (UK), and NASA Goddard Space Flight Center (USA). AIA data are courtesy of SDO (NASA) and the AIA consortium. The authors thank the open data policy of the WAVES/Wind instrument team. The GOES 14--15 X-ray data are produced in real time by the NOAA Space Weather Prediction Center (SWPC) and are distributed by the NOAA National Geophysical Data Center (NGDC).

%





\bibliography{ads}{}

\begin{thebibliography}{}
\expandafter\ifx\csname natexlab\endcsname\relax\def\natexlab#1{#1}\fi
\providecommand{\url}[1]{\href{#1}{#1}}
\providecommand{\dodoi}[1]{doi:~\href{http://doi.org/#1}{\nolinkurl{#1}}}
\providecommand{\doeprint}[1]{\href{http://ascl.net/#1}{\nolinkurl{http://ascl.net/#1}}}
\providecommand{\doarXiv}[1]{\href{https://arxiv.org/abs/#1}{\nolinkurl{https://arxiv.org/abs/#1}}}

\bibitem[{{Arnaud} \& {Raymond}(1992)}]{1992ApJ...398..394A}
{Arnaud}, M., \& {Raymond}, J. 1992, \apj, 398, 394, \dodoi{10.1086/171864}

\bibitem[{{Arnaud} \& {Rothenflug}(1985)}]{1985A&AS...60..425A}
{Arnaud}, M., \& {Rothenflug}, R. 1985, \aaps, 60, 425

\bibitem[{Bornmann {et~al.}(1996)Bornmann, Speich, Hirman, Matheson, Grubb,
  Garcia, \& Viereck}]{10.1117/12.254076}
Bornmann, P.~L., Speich, D., Hirman, J., {et~al.} 1996, in GOES-8 and Beyond,
  ed. E.~R. Washwell, Vol. 2812, International Society for Optics and Photonics
  (SPIE), 291 -- 298, \dodoi{10.1117/12.254076}

\bibitem[{{Bougeret} {et~al.}(1995){Bougeret}, {Kaiser}, {Kellogg}, {Manning},
  {Goetz}, {Monson}, {Monge}, {Friel}, {Meetre}, {Perche}, {Sitruk}, \&
  {Hoang}}]{1995SSRv...71..231B}
{Bougeret}, J.~L., {Kaiser}, M.~L., {Kellogg}, P.~J., {et~al.} 1995, \ssr, 71,
  231, \dodoi{10.1007/BF00751331}

\bibitem[{{Bu{\v{c}}{\'\i}k}(2020)}]{2020SSRv..216...24B}
{Bu{\v{c}}{\'\i}k}, R. 2020, \ssr, 216, 24, \dodoi{10.1007/s11214-020-00650-5}

\bibitem[{{Bu{\v{c}}{\'\i}k} {et~al.}(2014){Bu{\v{c}}{\'\i}k}, {Innes}, {Mall},
  {Korth}, {Mason}, \& {G{\'o}mez-Herrero}}]{2014ApJ...786...71B}
{Bu{\v{c}}{\'\i}k}, R., {Innes}, D.~E., {Mall}, U., {et~al.} 2014, Astrophys.
  J., 786, 71, \dodoi{10.1088/0004-637X/786/1/71}

\bibitem[{{Bu{\v{c}}{\'\i}k} {et~al.}(2016){Bu{\v{c}}{\'\i}k}, {Innes},
  {Mason}, \& {Wiedenbeck}}]{2016ApJ...833...63B}
{Bu{\v{c}}{\'\i}k}, R., {Innes}, D.~E., {Mason}, G.~M., \& {Wiedenbeck}, M.~E.
  2016, Astrophys. J., 833, 63, \dodoi{10.3847/1538-4357/833/1/63}

\bibitem[{{Bu{\v{c}}{\'\i}k} {et~al.}(2018{\natexlab{a}}){Bu{\v{c}}{\'\i}k},
  {Innes}, {Mason}, {Wiedenbeck}, {G{\'o}mez-Herrero}, \&
  {Nitta}}]{2018ApJ...852...76B}
{Bu{\v{c}}{\'\i}k}, R., {Innes}, D.~E., {Mason}, G.~M., {et~al.}
  2018{\natexlab{a}}, Astrophys. J., 852, 76, \dodoi{10.3847/1538-4357/aa9d8f}

\bibitem[{{Bu{\v{c}}{\'\i}k} {et~al.}(2018{\natexlab{b}}){Bu{\v{c}}{\'\i}k},
  {Wiedenbeck}, {Mason}, {G{\'o}mez-Herrero}, {Nitta}, \&
  {Wang}}]{2018ApJ...869L..21B}
{Bu{\v{c}}{\'\i}k}, R., {Wiedenbeck}, M.~E., {Mason}, G.~M., {et~al.}
  2018{\natexlab{b}}, Astrophys. J. Lett., 869, L21,
  \dodoi{10.3847/2041-8213/aaf37f}

\bibitem[{{Chen} {et~al.}(2013){Chen}, {Ip}, \& {Innes}}]{2013ApJ...769...96C}
{Chen}, N., {Ip}, W.-H., \& {Innes}, D. 2013, \apj, 769, 96,
  \dodoi{10.1088/0004-637X/769/2/96}

\bibitem[{{Chen} {et~al.}(2015){Chen}, {Bu{\v{c}}{\'\i}k}, {Innes}, \&
  {Mason}}]{2015AA...580A..16C}
{Chen}, N.-H., {Bu{\v{c}}{\'\i}k}, R., {Innes}, D.~E., \& {Mason}, G.~M. 2015,
  Astron. Astrophys., 580, A16, \dodoi{10.1051/0004-6361/201525618}

\bibitem[{{Cheung} {et~al.}(2015){Cheung}, {Boerner}, {Schrijver}, {Testa},
  {Chen}, {Peter}, \& {Malanushenko}}]{2015ApJ...807..143C}
{Cheung}, M. C.~M., {Boerner}, P., {Schrijver}, C.~J., {et~al.} 2015, \apj,
  807, 143, \dodoi{10.1088/0004-637X/807/2/143}

\bibitem[{{Cohen} {et~al.}(1999{\natexlab{a}}){Cohen}, {Mewaldt}, {Leske},
  {Cummings}, {Stone}, {Wiedenbeck}, {Christian}, \& {von
  Rosenvinge}}]{1999GeoRL..26.2697C}
{Cohen}, C.~M.~S., {Mewaldt}, R.~A., {Leske}, R.~A., {et~al.}
  1999{\natexlab{a}}, \grl, 26, 2697, \dodoi{10.1029/1999GL900560}

\bibitem[{{Cohen} {et~al.}(1999{\natexlab{b}}){Cohen}, {Cummings}, {Leske},
  {Mewaldt}, {Stone}, {Dougherty}, {Wiedenbeck}, {Christian}, \& {von
  Rosenvinge}}]{1999GeoRL..26..149C}
{Cohen}, C.~M.~S., {Cummings}, A.~C., {Leske}, R.~A., {et~al.}
  1999{\natexlab{b}}, \grl, 26, 149, \dodoi{10.1029/1998GL900218}

\bibitem[{{Dere} {et~al.}(2019){Dere}, {Del Zanna}, {Young}, {Landi}, \&
  {Sutherland}}]{2019ApJS..241...22D}
{Dere}, K.~P., {Del Zanna}, G., {Young}, P.~R., {Landi}, E., \& {Sutherland},
  R.~S. 2019, \apjs, 241, 22, \dodoi{10.3847/1538-4365/ab05cf}

\bibitem[{{DiFabio} {et~al.}(2008){DiFabio}, {Guo}, {M{\"o}bius}, {Klecker},
  {Kucharek}, {Mason}, \& {Popecki}}]{2008ApJ...687..623D}
{DiFabio}, R., {Guo}, Z., {M{\"o}bius}, E., {et~al.} 2008, Astrophys. J., 687,
  623, \dodoi{10.1086/591833}

\bibitem[{{Drake} {et~al.}(2009){Drake}, {Cassak}, {Shay}, {Swisdak}, \&
  {Quataert}}]{2009ApJ...700L..16D}
{Drake}, J.~F., {Cassak}, P.~A., {Shay}, M.~A., {Swisdak}, M., \& {Quataert},
  E. 2009, Astrophys. J. Lett., 700, L16, \dodoi{10.1088/0004-637X/700/1/L16}

\bibitem[{{Dwyer} {et~al.}(2001){Dwyer}, {Mason}, {Mazur}, {Gold}, {Krimigis},
  {M{\"o}bius}, \& {Popecki}}]{2001ApJ...563..403D}
{Dwyer}, J.~R., {Mason}, G.~M., {Mazur}, J.~E., {et~al.} 2001, \apj, 563, 403,
  \dodoi{10.1086/323692}

\bibitem[{{Feldman}(1992)}]{1992PhyS...46..202F}
{Feldman}, U. 1992, \physscr, 46, 202, \dodoi{10.1088/0031-8949/46/3/002}

\bibitem[{{Fu} {et~al.}(2020){Fu}, {Guo}, {Li}, \& {Li}}]{2020ApJ...890..161F}
{Fu}, X., {Guo}, F., {Li}, H., \& {Li}, X. 2020, \apj, 890, 161,
  \dodoi{10.3847/1538-4357/ab6d68}

\bibitem[{{Gloeckler} \& {Geiss}(1998)}]{1998SSRv...84..275G}
{Gloeckler}, G., \& {Geiss}, J. 1998, Space Sci. Rev., 84, 275

\bibitem[{{Grevesse} \& {Sauval}(1998)}]{1998SSRv...85..161G}
{Grevesse}, N., \& {Sauval}, A.~J. 1998, \ssr, 85, 161,
  \dodoi{10.1023/A:1005161325181}

\bibitem[{{Innes} {et~al.}(2016){Innes}, {Bu{\v{c}}{\'\i}k}, {Guo}, \&
  {Nitta}}]{2016AN....337.1024I}
{Innes}, D.~E., {Bu{\v{c}}{\'\i}k}, R., {Guo}, L.~J., \& {Nitta}, N. 2016,
  Astron. Nachr., 337, 1024, \dodoi{10.1002/asna.201612428}

\bibitem[{{Kartavykh} {et~al.}(2008){Kartavykh}, {Dr{\"o}ge}, {Klecker},
  {Kocharov}, {Kovaltsov}, \& {M{\"o}bius}}]{2008ApJ...681.1653K}
{Kartavykh}, Y.~Y., {Dr{\"o}ge}, W., {Klecker}, B., {et~al.} 2008, Astrophys.
  J., 681, 1653, \dodoi{10.1086/588649}

\bibitem[{{Kartavykh} {et~al.}(2020){Kartavykh}, {Dr{\"o}ge}, {Klecker},
  {Kovaltsov}, \& {Ostryakov}}]{2020ApJ...888...48K}
{Kartavykh}, Y.~Y., {Dr{\"o}ge}, W., {Klecker}, B., {Kovaltsov}, G.~A., \&
  {Ostryakov}, V.~M. 2020, Astrophys. J., 888, 48,
  \dodoi{10.3847/1538-4357/ab584e}

\bibitem[{{Klecker} {et~al.}(2007){Klecker}, {M{\"o}bius}, \&
  {Popecki}}]{2007SSRv..130..273K}
{Klecker}, B., {M{\"o}bius}, E., \& {Popecki}, M.~A. 2007, Space Sci. Rev.,
  130, 273, \dodoi{10.1007/s11214-007-9207-1}

\bibitem[{{Kocharov} {et~al.}(2000){Kocharov}, {Kovaltsov}, {Torsti}, \&
  {Ostryakov}}]{2000A&A...357..716K}
{Kocharov}, L., {Kovaltsov}, G.~A., {Torsti}, J., \& {Ostryakov}, V.~M. 2000,
  Astron. Astrophys., 357, 716

\bibitem[{{Kocharov} \& {Kocharov}(1984)}]{1984SSRv...38...89K}
{Kocharov}, L.~G., \& {Kocharov}, G.~E. 1984, Space Sci. Rev., 38, 89,
  \dodoi{10.1007/BF00180337}

\bibitem[{{Kovaltsov} {et~al.}(2001){Kovaltsov}, {Barghouty}, {Kocharov},
  {Ostryakov}, \& {Torsti}}]{2001A&A...375.1075K}
{Kovaltsov}, G.~A., {Barghouty}, A.~F., {Kocharov}, L., {Ostryakov}, V.~M., \&
  {Torsti}, J. 2001, Astron. Astrophys., 375, 1075,
  \dodoi{10.1051/0004-6361:20010877}

\bibitem[{{Krucker} {et~al.}(1999){Krucker}, {Larson}, {Lin}, \&
  {Thompson}}]{1999ApJ...519..864K}
{Krucker}, S., {Larson}, D.~E., {Lin}, R.~P., \& {Thompson}, B.~J. 1999, \apj,
  519, 864, \dodoi{10.1086/307415}

\bibitem[{{Kumar} {et~al.}(2017){Kumar}, {Eichler}, {Gaspari}, \&
  {Spitkovsky}}]{2017ApJ...835..295K}
{Kumar}, R., {Eichler}, D., {Gaspari}, M., \& {Spitkovsky}, A. 2017, Astrophys.
  J., 835, 295, \dodoi{10.3847/1538-4357/835/2/295}

\bibitem[{{Landi} {et~al.}(2002){Landi}, {Feldman}, \&
  {Dere}}]{2002ApJS..139..281L}
{Landi}, E., {Feldman}, U., \& {Dere}, K.~P. 2002, \apjs, 139, 281,
  \dodoi{10.1086/337949}

\bibitem[{{Lemen} {et~al.}(2012){Lemen}, {Title}, {Akin}, {Boerner}, {Chou},
  {Drake}, {Duncan}, {Edwards}, {Friedlaender}, {Heyman}, {Hurlburt}, {Katz},
  {Kushner}, {Levay}, {Lindgren}, {Mathur}, {McFeaters}, {Mitchell}, {Rehse},
  {Schrijver}, {Springer}, {Stern}, {Tarbell}, {Wuelser}, {Wolfson}, {Yanari},
  {Bookbinder}, {Cheimets}, {Caldwell}, {Deluca}, {Gates}, {Golub}, {Park},
  {Podgorski}, {Bush}, {Scherrer}, {Gummin}, {Smith}, {Auker}, {Jerram},
  {Pool}, {Soufli}, {Windt}, {Beardsley}, {Clapp}, {Lang}, \&
  {Waltham}}]{2012SoPh..275...17L}
{Lemen}, J.~R., {Title}, A.~M., {Akin}, D.~J., {et~al.} 2012, \solphys, 275,
  17, \dodoi{10.1007/s11207-011-9776-8}

\bibitem[{{Liu} {et~al.}(2006){Liu}, {Petrosian}, \&
  {Mason}}]{2006ApJ...636..462L}
{Liu}, S., {Petrosian}, V., \& {Mason}, G.~M. 2006, Astrophys. J., 636, 462,
  \dodoi{10.1086/497883}

\bibitem[{{Luhn} {et~al.}(1987){Luhn}, {Klecker}, {Hovestadt}, \&
  {Moebius}}]{1987ApJ...317..951L}
{Luhn}, A., {Klecker}, B., {Hovestadt}, D., \& {Moebius}, E. 1987, \apj, 317,
  951, \dodoi{10.1086/165343}

\bibitem[{{Mann} {et~al.}(1999){Mann}, {Jansen}, {MacDowall}, {Kaiser}, \&
  {Stone}}]{1999A&A...348..614M}
{Mann}, G., {Jansen}, F., {MacDowall}, R.~J., {Kaiser}, M.~L., \& {Stone},
  R.~G. 1999, \aap, 348, 614

\bibitem[{{Mason}(2007)}]{2007SSRv..130..231M}
{Mason}, G.~M. 2007, Space Sci. Rev., 130, 231,
  \dodoi{10.1007/s11214-007-9156-8}

\bibitem[{{Mason} {et~al.}(2000){Mason}, {Dwyer}, \&
  {Mazur}}]{2000ApJ...545L.157M}
{Mason}, G.~M., {Dwyer}, J.~R., \& {Mazur}, J.~E. 2000, Astrophys. J. Lett.,
  545, L157, \dodoi{10.1086/317886}

\bibitem[{{Mason} \& {Klecker}(2018)}]{2018ApJ...862....7M}
{Mason}, G.~M., \& {Klecker}, B. 2018, Astrophys. J., 862, 7,
  \dodoi{10.3847/1538-4357/aac94c}

\bibitem[{{Mason} {et~al.}(2002){Mason}, {Mazur}, \&
  {Dwyer}}]{2002ApJ...565L..51M}
{Mason}, G.~M., {Mazur}, J.~E., \& {Dwyer}, J.~R. 2002, Astrophys. J. Lett.,
  565, L51, \dodoi{10.1086/339135}

\bibitem[{{Mason} {et~al.}(2004){Mason}, {Mazur}, {Dwyer}, {Jokipii}, {Gold},
  \& {Krimigis}}]{2004ApJ...606..555M}
{Mason}, G.~M., {Mazur}, J.~E., {Dwyer}, J.~R., {et~al.} 2004, Astrophys. J.,
  606, 555, \dodoi{10.1086/382864}

\bibitem[{{Mason} {et~al.}(1994){Mason}, {Mazur}, \&
  {Halmilton}}]{1994ApJ...425..843M}
{Mason}, G.~M., {Mazur}, J.~E., \& {Halmilton}, D.~C. 1994, \apj, 425, 843,
  \dodoi{10.1086/174029}

\bibitem[{{Mason} {et~al.}(2009){Mason}, {Nitta}, {Cohen}, \&
  {Wiedenbeck}}]{2009ApJ...700L..56M}
{Mason}, G.~M., {Nitta}, N.~V., {Cohen}, C.~M.~S., \& {Wiedenbeck}, M.~E. 2009,
  Astrophys. J. Lett., 700, L56, \dodoi{10.1088/0004-637X/700/1/L56}

\bibitem[{{Mason} {et~al.}(2016){Mason}, {Nitta}, {Wiedenbeck}, \&
  {Innes}}]{2016ApJ...823..138M}
{Mason}, G.~M., {Nitta}, N.~V., {Wiedenbeck}, M.~E., \& {Innes}, D.~E. 2016,
  Astrophys. J., 823, 138, \dodoi{10.3847/0004-637X/823/2/138}

\bibitem[{{Mason} {et~al.}(1998){Mason}, {Gold}, {Krimigis}, {Mazur},
  {Andrews}, {Daley}, {Dwyer}, {Heuerman}, {James}, {Kennedy}, {Lefevere},
  {Malcolm}, {Tossman}, \& {Walpole}}]{1998SSRv...86..409M}
{Mason}, G.~M., {Gold}, R.~E., {Krimigis}, S.~M., {et~al.} 1998, \ssr, 86, 409,
  \dodoi{10.1023/A:1005079930780}

\bibitem[{{Mazzotta} {et~al.}(1998){Mazzotta}, {Mazzitelli}, {Colafrancesco},
  \& {Vittorio}}]{1998A&AS..133..403M}
{Mazzotta}, P., {Mazzitelli}, G., {Colafrancesco}, S., \& {Vittorio}, N. 1998,
  \aaps, 133, 403, \dodoi{10.1051/aas:1998330}

\bibitem[{{Miller}(1998)}]{1998SSRv...86...79M}
{Miller}, J.~A. 1998, Space Sci. Rev., 86, 79, \dodoi{10.1023/A:1005066209536}

\bibitem[{{M{\"o}bius} {et~al.}(2000){M{\"o}bius}, {Klecker}, {Popecki},
  {Morris}, {Mason}, {Stone}, {Bogdanov}, {Dwyer}, {Galvin}, {Heirtzler},
  {Hovestadt}, {Kistler}, \& {Siren}}]{2000AIPC..528..131M}
{M{\"o}bius}, E., {Klecker}, B., {Popecki}, M.~A., {et~al.} 2000, in American
  Institute of Physics Conference Series, Vol. 528, Acceleration and Transport
  of Energetic Particles Observed in the Heliosphere, ed. R.~A. {Mewaldt},
  J.~R. {Jokipii}, M.~A. {Lee}, E.~{M{\"o}bius}, \& T.~H. {Zurbuchen},
  131--134, \dodoi{10.1063/1.1324296}

\bibitem[{{Morgan} \& {Druckm{\"u}ller}(2014)}]{2014SoPh..289.2945M}
{Morgan}, H., \& {Druckm{\"u}ller}, M. 2014, \solphys, 289, 2945,
  \dodoi{10.1007/s11207-014-0523-9}

\bibitem[{{Mulay} {et~al.}(2019){Mulay}, {Sharma}, {Valori}, {V{\'a}squez},
  {Del Zanna}, {Mason}, \& {Oberoi}}]{2019A&A...632A.108M}
{Mulay}, S.~M., {Sharma}, R., {Valori}, G., {et~al.} 2019, \aap, 632, A108,
  \dodoi{10.1051/0004-6361/201936369}

\bibitem[{{Mulay} {et~al.}(2016){Mulay}, {Tripathi}, {Del Zanna}, \&
  {Mason}}]{2016A&A...589A..79M}
{Mulay}, S.~M., {Tripathi}, D., {Del Zanna}, G., \& {Mason}, H. 2016, \aap,
  589, A79, \dodoi{10.1051/0004-6361/201527473}

\bibitem[{{Nitta} {et~al.}(2015){Nitta}, {Mason}, {Wang}, {Cohen}, \&
  {Wiedenbeck}}]{2015ApJ...806..235N}
{Nitta}, N.~V., {Mason}, G.~M., {Wang}, L., {Cohen}, C. M.~S., \& {Wiedenbeck},
  M.~E. 2015, Astrophys. J., 806, 235, \dodoi{10.1088/0004-637X/806/2/235}

\bibitem[{{Nitta} {et~al.}(2006){Nitta}, {Reames}, {De Rosa}, {Liu}, {Yashiro},
  \& {Gopalswamy}}]{2006ApJ...650..438N}
{Nitta}, N.~V., {Reames}, D.~V., {De Rosa}, M.~L., {et~al.} 2006, Astrophys.
  J., 650, 438, \dodoi{10.1086/507442}

\bibitem[{{Pesnell} {et~al.}(2012){Pesnell}, {Thompson}, \&
  {Chamberlin}}]{2012SoPh..275....3P}
{Pesnell}, W.~D., {Thompson}, B.~J., \& {Chamberlin}, P.~C. 2012, \solphys,
  275, 3, \dodoi{10.1007/s11207-011-9841-3}

\bibitem[{{Reames}(2019)}]{2019Atoms...7..104R}
{Reames}. 2019, Atoms, 7, 104, \dodoi{10.3390/atoms7040104}

\bibitem[{{Reames}(1988)}]{1988ApJ...325L..53R}
{Reames}, D.~V. 1988, Astrophys. J. Lett., 325, L53, \dodoi{10.1086/185109}

\bibitem[{{Reames}(1990)}]{1990ApJS...73..235R}
---. 1990, Astrophys. J. Suppl. Ser., 73, 235, \dodoi{10.1086/191456}

\bibitem[{{Reames}(1995)}]{1995AdSpR..15g..41R}
---. 1995, Advances in Space Research, 15, 41

\bibitem[{{Reames}(2002)}]{2002ApJ...571L..63R}
---. 2002, Astrophys. J. Lett., 571, L63, \dodoi{10.1086/341149}

\bibitem[{{Reames}(2013)}]{2013SSRv..175...53R}
---. 2013, \ssr, 175, 53, \dodoi{10.1007/s11214-013-9958-9}

\bibitem[{{Reames}(2017)}]{2017LNP...932.....R}
---. 2017, {Solar Energetic Particles}, Vol. 932 (Springer, Berlin),
  \dodoi{10.1007/978-3-319-50871-9}

\bibitem[{{Reames}(2018)}]{2018SSRv..214...61R}
---. 2018, \ssr, 214, 61, \dodoi{10.1007/s11214-018-0495-4}

\bibitem[{{Reames} {et~al.}(2014{\natexlab{a}}){Reames}, {Cliver}, \&
  {Kahler}}]{2014SoPh..289.3817R}
{Reames}, D.~V., {Cliver}, E.~W., \& {Kahler}, S.~W. 2014{\natexlab{a}}, Sol.
  Phys., 289, 3817, \dodoi{10.1007/s11207-014-0547-1}

\bibitem[{{Reames} {et~al.}(2014{\natexlab{b}}){Reames}, {Cliver}, \&
  {Kahler}}]{2014SoPh..289.4675R}
---. 2014{\natexlab{b}}, Sol. Phys., 289, 4675,
  \dodoi{10.1007/s11207-014-0589-4}

\bibitem[{{Reames} {et~al.}(2015){Reames}, {Cliver}, \&
  {Kahler}}]{2015SoPh..290.1761R}
---. 2015, Sol. Phys., 290, 1761, \dodoi{10.1007/s11207-015-0711-2}

\bibitem[{{Reames} {et~al.}(1988){Reames}, {Dennis}, {Stone}, \&
  {Lin}}]{1988ApJ...327..998R}
{Reames}, D.~V., {Dennis}, B.~R., {Stone}, R.~G., \& {Lin}, R.~P. 1988,
  Astrophys. J., 327, 998, \dodoi{10.1086/166257}

\bibitem[{{Reames} {et~al.}(1994){Reames}, {Meyer}, \& {von
  Rosenvinge}}]{1994ApJS...90..649R}
{Reames}, D.~V., {Meyer}, J.~P., \& {von Rosenvinge}, T.~T. 1994, Astrophys. J.
  Suppl. Ser., 90, 649, \dodoi{10.1086/191887}

\bibitem[{{Reames} \& {Ng}(2004)}]{2004ApJ...610..510R}
{Reames}, D.~V., \& {Ng}, C.~K. 2004, Astrophys. J., 610, 510,
  \dodoi{10.1086/421518}

\bibitem[{{Reames} \& {Stone}(1986)}]{1986ApJ...308..902R}
{Reames}, D.~V., \& {Stone}, R.~G. 1986, Astrophys. J., 308, 902,
  \dodoi{10.1086/164560}

\bibitem[{{Reames} {et~al.}(1985){Reames}, {von Rosenvinge}, \&
  {Lin}}]{1985ApJ...292..716R}
{Reames}, D.~V., {von Rosenvinge}, T.~T., \& {Lin}, R.~P. 1985, Astrophys. J.,
  292, 716, \dodoi{10.1086/163203}

\bibitem[{{Roth} \& {Temerin}(1997)}]{1997ApJ...477..940R}
{Roth}, I., \& {Temerin}, M. 1997, \apj, 477, 940, \dodoi{10.1086/303731}

\bibitem[{{Stone} {et~al.}(1998){Stone}, {Frandsen}, {Mewaldt}, {Christian},
  {Margolies}, {Ormes}, \& {Snow}}]{1998SSRv...86....1S}
{Stone}, E.~C., {Frandsen}, A.~M., {Mewaldt}, R.~A., {et~al.} 1998, \ssr, 86,
  1, \dodoi{10.1023/A:1005082526237}

\bibitem[{{Temerin} \& {Roth}(1992)}]{1992ApJ...391L.105T}
{Temerin}, M., \& {Roth}, I. 1992, Astrophys. J. Lett., 391, L105,
  \dodoi{10.1086/186408}

\bibitem[{{Wang} {et~al.}(2016){Wang}, {Krucker}, {Mason}, {Lin}, \&
  {Li}}]{2016A&A...585A.119W}
{Wang}, L., {Krucker}, S., {Mason}, G.~M., {Lin}, R.~P., \& {Li}, G. 2016,
  Astron. Astrophys., 585, A119, \dodoi{10.1051/0004-6361/201527270}

\bibitem[{{Wang} {et~al.}(2012){Wang}, {Lin}, {Krucker}, \&
  {Mason}}]{2012ApJ...759...69W}
{Wang}, L., {Lin}, R.~P., {Krucker}, S., \& {Mason}, G.~M. 2012, Astrophys. J.,
  759, 69, \dodoi{10.1088/0004-637X/759/1/69}

\bibitem[{{Wang} {et~al.}(2006){Wang}, {Pick}, \&
  {Mason}}]{2006ApJ...639..495W}
{Wang}, Y.~M., {Pick}, M., \& {Mason}, G.~M. 2006, Astrophys. J., 639, 495,
  \dodoi{10.1086/499355}

\bibitem[{{Wiedenbeck} {et~al.}(2010){Wiedenbeck}, {Cohen}, {Leske}, {Mewaldt},
  {Cummings}, {Stone}, \& {von Rosenvinge}}]{2010ApJ...719.1212W}
{Wiedenbeck}, M.~E., {Cohen}, C.~M.~S., {Leske}, R.~A., {et~al.} 2010, \apj,
  719, 1212, \dodoi{10.1088/0004-637X/719/2/1212}

\end{thebibliography}
\bibliographystyle{aasjournal}



\end{document}